# Mechanism of climb in dislocation-nanovoid interaction

A. Dutta[a], M. Bhattacharya[b,*], N. Gayathri[b], G. C. Das[a], and P. Barat[b]

[a]Department of Metallurgical and Materials Engineering, Jadavpur University, Kolkata 700 032, India

[b]Variable Energy Cyclotron Centre, 1/AF, Bidhannagar, Kolkata 700 064, India

**Abstract**

Techniques of atomistic simulation have been employed to unfold the mechanism of a newly observed process of climb of a dislocation line during its depinning at a nanosized void. The analyses prove that in contrast to the conventional notion of dislocation climb, this unique process is not a diffusion-controlled phenomenon. Instead, the gradient of structural energy of the system alone suffices to manifest in the climb motion. The energetics of this process has been investigated through molecular statics and dynamics computations and the causes of discrepancies are discussed. In order to further explore these concepts in a quantitative manner, we have developed a robust simulation strategy to accurately measure the reduction in energy of the nanovoid during dislocation climb. The results explain the lowering of critical depinning load and effect of thermal assistance to void-induced climb. Moreover, we highlight the key role played by the curvature of dislocation line in mediating this mechanism.

*Corresponding author:

Dr. Mishreyee Bhattacharya
Phone : +91-33-2318-2411, Fax : +91-33-2334-6871
Email: mishreyee@vecc.gov.in

# 1. Introduction

Climb of dislocations is a crucial phenomenon deciding the long term durability of many structural materials, particularly in extreme conditions of load, temperature and radiation. This is owing to the fact that in contrast to dislocation glide, climb takes place on a much larger time scale. This nonconservative motion is usually perceived as a slow process, taking place via the complex mechanism of lattice and pipe diffusions of point defects [1-4]. Since the concentrations of point defects get altered during the process of radiation damage, the diffusion based climb is well known to be affected by the irradiation of materials [5,6]. However, besides varying the point defect concentrations, irradiation also causes the nucleation and growth of voids [7-9], which act as strong pinning obstacles. Numerous studies have been carried out to determine the nature of dislocation pinning at voids, in the context of their influence on macroscopic plastic properties of the metallic materials. In particular, a rigorous insight of the dislocation-void interaction has been gained on account of many atomistic simulations performed over the last few years. Several of these simulations [10-15] have shown that if the applied shear load is large enough to cause depinning, a part of the dislocation segment climbs while passing through a nanosized void. In such cases, the nanovoid acts a source of vacancies, whereas the dislocation becomes a sink. This void-induced climb (VIC) follows, in effect, a mechanism different from the one in which the extent of irradiation governs the rate of climb by controlling the diffusion flux of point defects towards the dislocation.

Despite being a fundamentally unique mechanism, the phenomenon of VIC has not been investigated in requisite detail, for most of the related studies were primarily aimed at measuring the pinning strength of nanovoids with variation in void radius, size of the simulation cell and temperature etc. Nevertheless, a close look at the observations emerging out of the earlier atomistic simulations can reveal certain contradictions owing to this lack of understanding of VIC. For instance, the occurrence of this process has been seen in the molecular statics (MS) simulations [10,12]. Since MS computations involve a relaxation algorithm which only drives the system towards a local energy minimum, one would expect the absence of any potential energy barrier to VIC and the thermal activation should not be relevant to this phenomenon. However, in their molecular dynamics (MD) simulations Hatano and Matsui [16] observed that the temperature dependence of the critical depinning load was linked to the phenomenon of VIC, thereby giving an explanation for the absence of temperature sensitivity to dislocation depinning at nanovoid in fcc copper, where the VIC was not observed. They also extended this concept to further explain the more prominent temperature dependence in larger nanovoids, where the extent of VIC is more than in the smaller voids. This notion of linking thermal assistance to the occurrence of VIC is apparently contradictive to the expectation of thermal insensitivity drawn from the observations of MS simulations. Furthermore, Osetsky *et al*. [13] have reported that the occurrence of VIC is associated with some reduction in critical depinning load. Thus, besides uncovering the significance of temperature in VIC, a comprehensive understanding of this phenomenon must also offer an explanation for the drop in the critical depinning load.

In this work, we explicitly focus on the phenomenon of VIC by analyzing this process using both MS and MD computations. Recently, Monnet [17] has demonstrated that different stages of a dynamical process can be identified in its overall energetics and applied this idea to recognize the steps involved in the pinning-depinning process of a dislocation at nanovoid. In this study by Monnet, six distinct regimes of the process (Fig. 2 in Ref. [17]) could be identified and the fifth stage depicting the energy barrier appertains to the actual process of depinning where the dislocation line starts bowing out. Nonetheless, various effects like elastic interaction between dislocation and the nanovoid, self interaction between the bowing out dislocation segments, slip-induced formation of surface steps and the VIC get compounded in this single stage. Consequently, it becomes a challenging task to decouple the energy of VIC from the energetics of the whole depinning regime. The present work adopts a similar concept as done by Monnet, but with a different technique of implementation where in lieu of a straightforward analysis of the energetics, we express the measurements with respect to a hypothetical reference structure created using a novel simulation scheme as detailed in a forthcoming section. Here, we focus on the energetics of VIC and attempt to explore the underlying mechanism. In the case of dislocation-nanovoid system, we can write the change in potential energy due to VIC as $\Delta E = E_{climb} + \Delta E_{void}$, where $E_{climb}$ is the energy change associated with the climb of the dislocation and $\Delta E_{void}$ denotes the change in energy of the nanovoid due to emission of vacancies. It is worthwhile to point out that although the whole pinning-depinning transition involves several ingredients of its energetics as stated above, only $E_{climb}$ and $\Delta E_{void}$ remain the dominating factors in the context of VIC, for the other factors are also present in cases of depinning without VIC. As the MS algorithm always relaxes the system to a structure with lower energy than the initial structure, the final configuration with a climbed dislocation segment cannot be attained in the static simulation unless $\Delta E_{void} < -E_{climb} < 0$ as $E_{climb} > 0$ [2]. This criterion will be looked upon while discussing the outcome of the simulations.

We first provide an overview of the overall energetics of the process during pinning-depinning transition before resolving it further into the energetics associated with the dislocation and the obstacle separately. In section 2, we present the simulation schemes and results obtained from the MS and MD computations respectively, with explanation for the observations. Section 3 describes the hindrances against estimating the nanovoid's shrinkage energy ($\Delta E_{void}$) directly from the atomistic simulations and outlines an elegant method for surmounting those difficulties. The energy of dislocation climb ($E_{climb}$) is also calculated in order to provide the complete picture of the whole energetics. In section 4, the energetics is analyzed in detail and the role of dislocation line profile is highlighted in the context of thermal assistance to VIC. In addition, this section also discusses the structure of dislocation near the point of its contact with the nanovoid. Section 5 is devoted to explaining how the process of VIC subsides the critical depinning load. The results and conclusions are finally summed up in section 6.

## 2. Energetics of VIC: MS vs. MD

As the dissociated cores energetically prohibit the VIC in fcc crystals [16], we perform our simulations for edge dislocations in bcc solids. The MD++ [18] molecular dynamics code has been employed for the purpose of our atomistic simulations. We have observed the occurrence of VIC in iron (Fe), tungsten (W) and molybdenum (Mo). Fe and W exhibit much larger tendencies of climb which manifest in multiple jog steps and superjogs. Therefore, we select Mo for a detailed analysis as it produces a simpler structure having a climbed segment with the height of only single atomic plane, which makes it most suitable for studying the mechanism of VIC at a fundamental level.

*2.1 Molecular static simulation*

In essence, the MS simulation involves the relaxation of a given initial structure to a state of mechanical equilibrium in the presence of interatomic and external forces. In our computations, the Mo simulation cell with dimensions $\mathbf{C}_1 = 59.5a\langle 111\rangle/2$, $\mathbf{C}_2 = 20a\langle \bar{1}01\rangle$ and $\mathbf{C}_3 = 15a\langle 1\bar{2}1\rangle$ (lattice constant $a$ = 0.31472 nm) is constructed using the Finnis-Sinclair interatomic potential [19,20], later modified by Ackland and Thetford [21]. The edge dislocation with Burgers vector $a\langle 111\rangle/2$ in the glide plane $(\bar{1}01)$ is created by splicing two crystals with a dimension differing by one Burgers vector (see section 2.3.2 of Ref. [12] and chapter 4 of Ref. [22] for example), followed by energy minimization [10,12,22] using the conjugate gradient relaxation method [22]. The nanovoid, with its center on the glide plane of the edge dislocation, is created 3.5 nm away from the dislocation line by removing all the atoms within its radius and subsequently relaxing the system to obtain the initial static equilibrium. Periodic boundary condition (PBC) is imposed along the directions of the dislocation line and the Burgers vector, so that a periodic array of dislocations [12] is obtained in a slab of finite thickness. The boundary vector normal to the glide plane is tilted and thus the incremental shear strain is added in steps of $2\times 10^{-5}$. After each step, the system is relaxed to its minimum energy configuration. Consequently, the dislocation line keeps bowing out in order to maintain its mechanically equilibrated shape at each step of shear strain and the energy of the system is recorded until depinning takes place. On the whole, the pinning-depinning transition is simulated in a discrete quasistatic manner representative of a hypothetical process at absolute zero temperature.

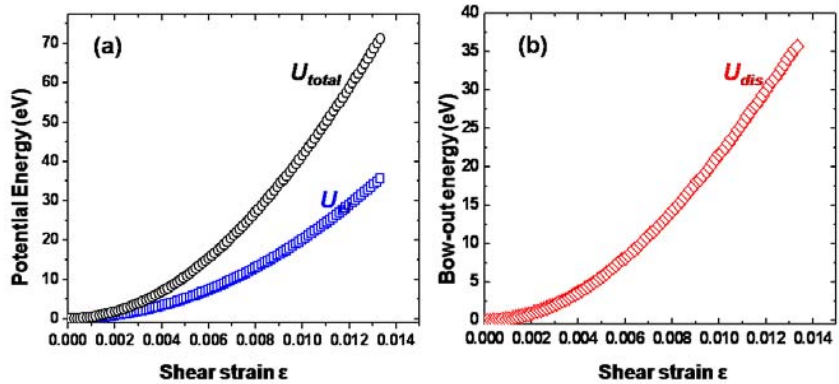

Fig. 1. (a) Rise in total potential energy of the system ($U_{tot}$) and elastic energy stored in the crystal ($U_{el}$) with increase in shear strain ($\varepsilon$) in Mo. (b) The bow-out energy of the curved dislocation line for the 2 nm

diameter nanovoid obtained by deducting the elastic energy from the total potential energy. The energies before the pinning and after the depinning (at $\varepsilon = 0.01333$) are omitted for the sake of brevity.

Interestingly, in the Mo crystal, the dislocation passes without undergoing any climb if the void diameter is less than 1.2 nm, whereas the VIC is always observed if the dislocation passes through a larger nanovoid. The rise in total energy of the system ($U_{total}(\varepsilon)$) with increasing shear strain ($\varepsilon$) is obtained directly from the simulation outcome. This energy can be resolved into two parts [13]: first, the crystal gains the elastic strain energy due to the internal shear stress and second, a part of the work done on the system is stored in the bowing out dislocation line. The average shear stress is given by $\tau(\varepsilon) = \partial U_{total}/\partial \varepsilon$, and the elastic energy stored in the crystal is computed as $U_{el} = \tau^2 V/(2\mu)$, where $V$ is the volume of the simulation cell and $\mu = 137.15$ GPa is the shear modulus of the Mo crystal. The bow-out energy of the dislocation is then obtained as $U_{dis} = U_{total} - U_{el}$. Figures 1a and b exhibit the elastic energy stored in the crystal and the bow-out energy of the dislocation line respectively, for a 2 nm diameter nanovoid in Mo as a pinning obstacle, where the VIC takes place by transfer of seven vacancies from the nanovoid to dislocation core. They indicate monotonic rise of both these energies in between the states of pinning and depinning, similar to the earlier results [10,13,17] for Fe.

2.2 *Molecular dynamics simulation*

Simulation setup for the MD computations is similar to that of the MS. After constructing the dislocation-nanovoid system, it is thermally equilibrated [22] at 50 K and maintained at that temperature using the Nosé-Hoover thermostat [23,24]. The dislocation is made to glide under the shear load which is created by applying traction forces on the top and bottom surfaces of the Mo slab and the time evolution of potential energy is obtained. As the phenomenon of VIC has been suggested to be thermally assisted at elevated temperatures [12,14,16], we purposefully select a low temperature in order to minimize the role of thermal activation in the present simulations. This also ensures minimum thermal fluctuations, thereby providing consistent results over repeated simulation runs.

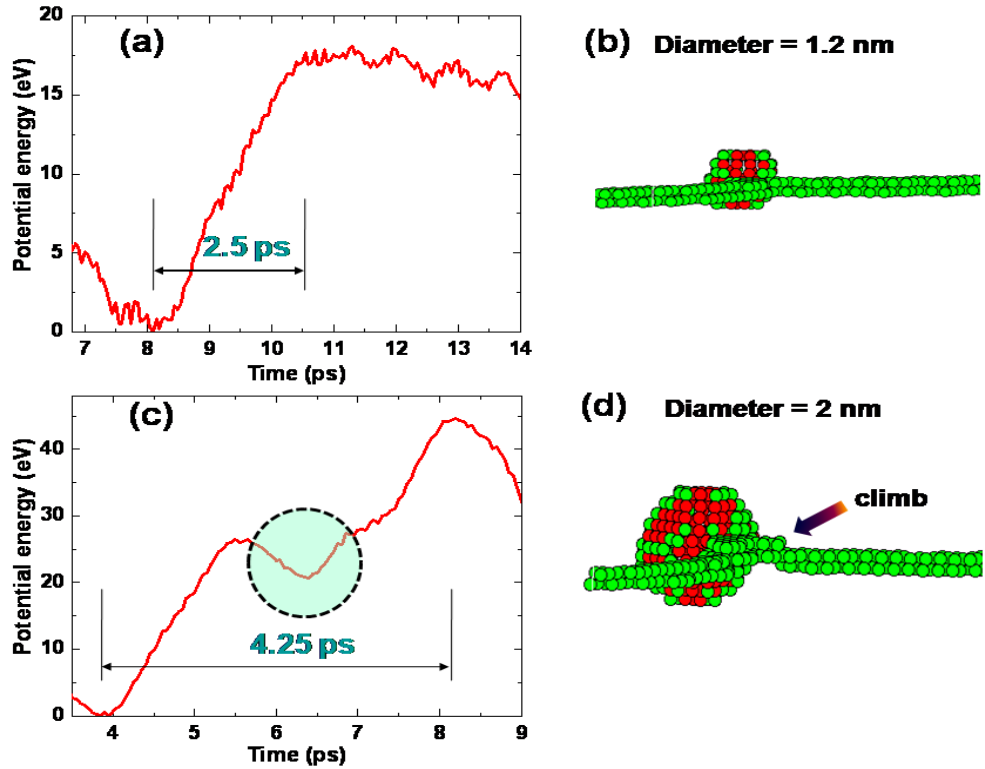

Fig. 2. (a) Part of the potential energy vs. time profile covering the pinning-depinning duration for the 1.2 nm diameter nanovoid at the critical depinning load of ~160 MPa and (b) the corresponding MD snapshot. (c) Energetics of the system with the 2 nm diameter nanovoid showing an intermediate drop in potential energy at the corresponding critical load of ~565 MPa and (d) the simulation snapshot where VIC is clearly observed. The potential energy values are shown with suitable offsets so that minima of the energy plots correspond to zero energy.

Even though the MD simulations also reveal the occurrence of VIC only above the nanovoid diameter of 1.2 nm, the trend of its energetics is qualitatively different from the MS result. Figure 2a shows the typical potential energy profile of the pinning-depinning process in MD simulation at the 1.2 nm nanovoid, where the obstacle fails to induce any dislocation climb (Fig. 2b). Here the energy keeps increasing consistently between the stages of pinning and depinning as marked in Fig. 2a. On the contrary, the energy profile of the system with the 2 nm void (Fig. 2c) indicates an intermediate drop in the otherwise inceasing energy during the pinning-depinning transition, where the dislocation line breaks away after absorbing seven vacancies from the nanovoid as displayed in Fig. 2d. These findings hint at a possible connection between the VIC and the observation of intermediate drop in the energy profile.

*2.3 Analysis of the simulation outcome*

We can regard the change in potential energy of the nanovoid ($E_{void}$) as the governing factor for the climb motion. Earlier, Knight and Burton [25] have indicated that a nanovoid has its inherent surface tension due to the free inner surface. This has been held responsible for the tendency of a nanovoid to shrink by losing its vacancies to the dislocation. Nevertheless, the nature of climb investigated by Knight and Burton [25] involves the pipe diffusion of vacancies along the core of a climbing dislocation, which suddenly encounters a void. This diffusion mechanism, albeit enhanced by the surface tension of the void, is essentially thermally activated whereas the observation of VIC in the present and previous MS ($T = 0$ K) simulations prove a non-diffusive route of the process. Even in the MD computations, the entire pinning-depinning transition takes place within a few picoseconds (Fig. 2c), which is fast enough to rule out the possibility of diffusive kinetics at such a low temperature ($T = 50$ K).

The above discussion shows that the energetics of VIC corresponds to the change in structural energy of the system during dislocation climb and in the absence of vacancy diffusion in both MS and low temperature MD, one can expect similar trends of energetics in the MS and MD results. However, in spite of the fact that the process involves the transfer of same number of vacancies in both MS and MD, we have already seen that Fig. 1 does not show any distinct signature of VIC as seen in Fig. 2c. The reason why we miss the intermediate drop in the MS simulation leads us to point out that unlike the MD, MS does not consider time as an intrinsic parameter as the momentum vectors are absent in its basic algorithm. As a result, the actual dynamics of the system in between two consecutive relaxed states is not available and accordingly, any particular feature of the energetics associated with that dynamics is also lost. Furthermore, in the static computation, the change in the potential energy during the shrinkage of the nanovoid is hidden behind the large elastic energy stored in the crystal. On the other hand, in dynamic simulations, the dislocation line is always gliding as the applied shear load exceeds the critical depinning stress and the system is maintained at mechanical non-equilibrium. This prevents the crystal from gaining the same elastic energy as in MS and we are able to observe the intermediate drop in the energy profile. The present study demonstrates that the thermal vibration of atoms is not the only factor which can distinguish a static atomistic simulation from its dynamic counterpart. We also conclude that although the analysis of energetics obtained from MS outcome can successfully reveal the different stages associated with a process [17], the MD results must also be invested to gather the complete picture of its dynamics.

## 3. Energies of dislocation climb, dislocation line and the void shrinkage

The process of VIC takes place through transfer of vacancies from the nanovoid to dislocation, which results in modification of energies of the nanovoid as well as the dislocation. In this section we describe the methods of calculating these energies separately. We also explain the technique of estimating the line tension of dislocation, a quantity, which will be used during the discussion in section 5.

*3.1 Estimation of climb energy ($E_{climb}$)*

The change in the energy of the dislocation line due to formation of jog pair is calculated as function of the length of climbed segment. For this purpose, a single edge dislocation is introduced in the Mo slab of dimensions and crystal directions same as that used for the simulations of VIC as described in the previous section. The total potential energy of the simulation cell ($E_{cell}$) is recorded along with the total number of atoms (*N*) in the cell. Next, we create a similar cell with a portion of the dislocation line climbed up with the addition of $N_v$ vacancies. This structure can be created by splicing two crystals back to back, such that the crystal at the front contains a dislocation line at one atomic plane higher than the other. By changing the lengths of the two segments such that the total length of the dislocation line remains constant, one can simulate the addition of any number of vacancies to the dislocation line. We again measure the potential energy $E'_{cell}$ of this simulation cell and express the energy of dislocation climb as

$$E_{climb}(N_v) = E'_{cell} - E_{cell} + N_v E_{coh}, \qquad (1)$$

where $E_{coh}$ is the cohesive energy of the metal, which is -6.82 eV/atom for the interatomic potential used in this study. It must be noted that for this computation, the PBC has been terminated along the direction of the dislocation line. This is done because in the original VIC simulation described above, the jogs nucleate at the surface of the nanovoid, which is a free boundary. During the process of depinning, as the point of contact of the pinned dislocation segment slides along the nanovoid's periphery, it makes gradually varying orientation with respect to the surface. However, the present data assumes the dislocation to terminate on a free flat surface while the length of the climbed segment is varied to evaluate the climb energy. This approximation does not invalidate the fundamental physical concept, for it still gives a reasonable estimate of $E_{climb}$ without altering the principle conclusions.

*3.2 Estimation of line tension ($E_{edge}$)*

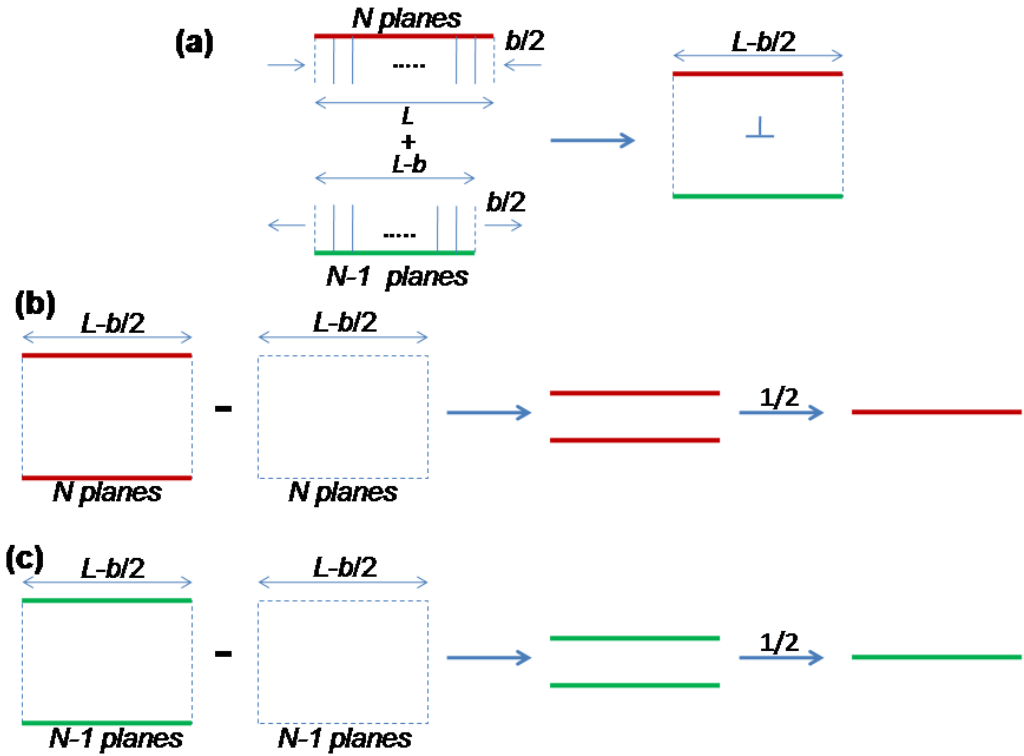

Fig. 3. (a) To create an edge dislocation, upper half-crystal containing the extra half plane is squeezed by half of the Burgers vector (*b*) and joined with the lower half-crystal, which has been expanded by the same extent. (b) For evaluation of energy of the top surface, potential energy of the crystal with full PBC and compressed by $b/2$, is subtracted from a similar crystal with free surfaces. The result is then halved to obtain the value for one surface only. (c) The energy of the bottom surface is computed in the same way as that of the top surface, but here the crystals are under tensile strain and have one atomic plane less than that in the previous case. In these schematics, the thin dotted lines represent the periodic boundaries, whereas the surfaces are denoted by the thicker solid lines.

The line tension of a dislocation is simply the energy of unit length of the dislocation line. Although the measurement of energy associated with a defect is a straightforward process in MD simulations, we cannot follow this direct approach for the defect energy would also include the energies of free top and bottom surfaces. At the same time, we cannot eliminate the free surfaces by applying PBC in the vertical direction as it would create negative dislocations at the interfaces of the simulation cell and contaminate the measurements. Moreover, the line tension in such a case would not correspond to the required value in a slab of finite thickness. Therefore, we measure the surface energies separately and deduct them from the total defect energy (including the energy of the dislocation line and the free surfaces) of the system to yield the dislocation line energy in the metallic slab. Noticeably, the energies of the top and bottom surfaces have to be computed independently since they have different numbers of atoms due to the extra half plane of the edge dislocation. Similarly, due consideration has to be given to the fact that the top surface is under compressive strain while the bottom surface has a tensile strain.

For the purpose of computing the energy of the free top surface, we first simulate a crystalline slab without any defect other than the free surfaces with crystal directions and dimension as in the VIC simulations, The number of atomic planes in the x direction must be equal to that in the upper half crystal of the VIC simulation cell and hence the crystal is at a compressive strain. The system is relaxed to minimum energy structure and the potential energy is evaluated. This energy corresponds to the energy of the strained surfaces and the elastic energy stored in the crystal. Now PBC is imposed along the vertical direction in order to remove the free surfaces, and the potential energy is measured again. In the absence of surfaces, this value belongs to the elastic energy of the compressed crystal only, which is deducted from the previous measurement to yield the surface energy. This is further halved to obtain the energy of the top surface only. In the same way, we can simulate a slab where the number of atomic planes is same as that in the lower half crystal of our VIC simulation. This slab is now under tensile strain and thus, we can again evaluate the energy of the bottom surface which is stretched in a tensile manner. Figure 3 shows a schematic which explains this methodology lucidly.

*3.3 Estimation of shrinkage energy: cut and erase method*

Emission of vacancies manifests in reduction in size of the nanovoid, accompanied by decrease in its energy. Evaluating the energy of void shrinkage proves to be a nontrivial task because expressing it as the straightforward difference between the energies of final (*post*-climb) and initial (*pre*-climb) configurations would not yield the correct result. This is on account of the fact that the simulation cell always contains the dislocation and hence, total energy of the cell also involves the energy of dislocation-nanovoid interaction, which is a function of the distance between the dislocation and nanovoid. As the continuum theories of dislocation-obstacle interactions break down at the atomic scale, calculation of this interaction energy by analytical means (Ref. [26] for instance) and assuming a seamless blending of elasticity theory and atomistic simulation would be a *naïve* and unreliable approach. Another issue in this regard is the plastic slip caused by the gliding dislocation, which manifests in a relative displacement of upper half of the crystal by one Burgers vector, with respect to the lower half. This creates protruding steps at the inner surface of the nanovoid [27], which raises the structural energy of the obstacle. These steps are formed as the dislocation enters and exits the obstacle and would be on the same glide plane had the breakaway occurred without VIC. In contrast, although the shrinkage of the nanovoid drives the phenomenon of VIC, the surface steps are still formed at the entry and exit, albeit on different glide planes due to the formation of jog pair during breakaway. Consequently, change in the energy of the nanovoid during shrinkage includes the energy of surface step as well. Typically, the energy of such steps is presumed to be much smaller as compared to that of the bowing out dislocation and is often ignored [17], yet it is large enough to be considered in the context of nanovoid's shrinkage. In order to filter out the shrinkage energy of nanovoid from the above mentioned sources of error, we have excogitated a novel simulation strategy capable of providing the desired value.

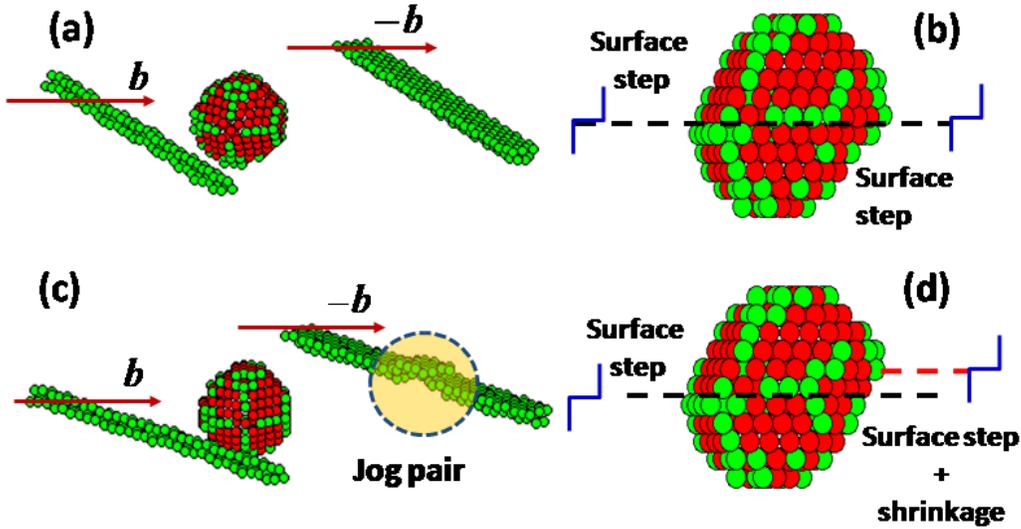

Fig. 4. (a) The starting configuration with the original dislocation on the left and negative dislocation on the right side of the nanovoid. (b) The final structure after relaxation. The surface steps on the nanovoid can be seen to lie on the same glide plane (dashed line). (c) The initial configuration where the negative dislocation now has a climbed segment. (d) After relaxation, the surface steps on the nanovoid are on different glide planes indicating its shrinkage.

The basic essence of this method lies in introducing another similar dislocation with opposite Burgers vector on other side of the nanovoid on the same glide plane, as demonstrated in Fig. 4a. As we relax the system with MS algorithm, both the dislocations approach the nanovoid due to the combined influence of dislocation-nanovoid attraction and dislocation-dislocation attraction. The dislocations must be placed at suitable distances from the nanovoid so that the elastic interactions are large enough to overcome the Peierls barrier. After relaxation, both the dislocations are found to intersect the nanovoid and annihilate each other. The remaining structure (Fig. 4b) is only a nanovoid of unaltered volume but with surface steps created by the dislocation dipole. In the same way, if the negative dislocation is made to host a jog pair (Fig. 4c), the simulation cell ends up with a shrunken nanovoid with surface steps as shown in Fig. 4d. Here the nanovoid loses as many vacancies ($N_v$) as the number of extra atoms used in creating the jog pair on the negative dislocation. A noticeable feature of this method is the flexibility of controlling the nanovoid's shrinkage by altering the size and position of the climbed segment on negative dislocation. Since the final structure is now devoid of any dislocation, we have overcome the issue of dislocation-nanovoid interaction. Similarly, we have been able to create a hypothetical structure of the nanovoid with surface steps but without any shrinkage (Fig. 4b). As the structures in both of Figs. 4b and d now have the surface steps, we can simply use the relation on the right hand side of Eq. (1), which in this case would yield the shrinkage energy $\Delta E_{void}$. A noteworthy feature of this scheme is that even if the energy of surface steps turns out to be many times larger than the shrinkage energy, it cannot affect the performance of the given method. As a matter of fact, the present results indicate the feasibility of explicitly computing the part of energetics responsible for VIC without exploring the energetics of the overall pinning-depinning dynamics. Instead, we simply hypothesize a case where the whole process proceeds in the same way, except the climb at the nanovoid. Deducting the energy of this hypothetical case from that of the actual process would cancel out the features common to their dynamics. Finally, only the energies of nanovoid's shrinkage and the dislocation climb remain as the dominating factors inducing the non-diffusive phenomenon of VIC. Nevertheless, in the view of possible errors due to improper relaxation of top and bottom surfaces of the slab, one can verify the results by repeating the measurements with full PBC thereby eliminating the free boundaries. In our study, removal of surfaces is found to cause only a negligible difference of 0.4%, which confirms the accuracy of the relaxation process. Although this method has been developed here with the aim of computing the shrinkage energy of nanovoid, the technique of cutting the obstacle and then erasing the dislocation is inherently a more generalized strategy, which can be used to measure the change in structural energy of any obstacle due to the passage of a dislocation through it.

## 4. Results and discussions

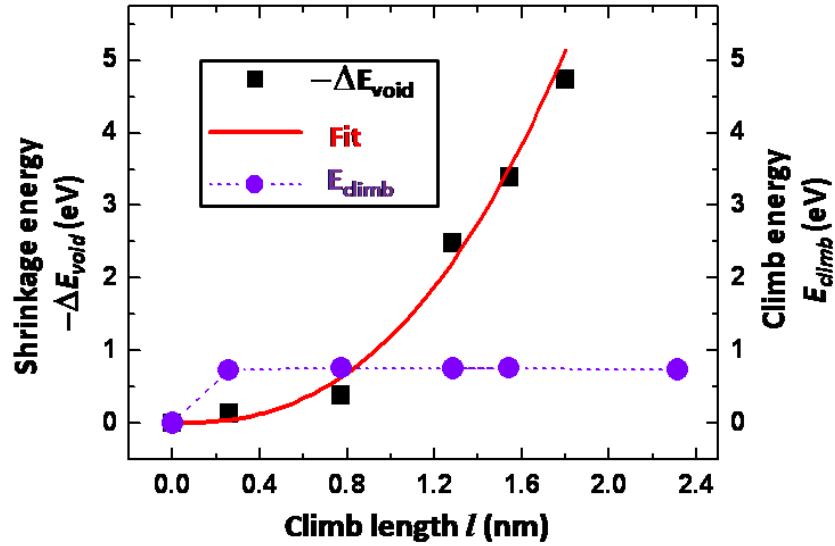

Fig. 5. Energy released due to the nanovoid's shrinkage ($\Delta E_{void}$) as a function of climb length ($l$) with the nonlinear fit and the jog formation energy ($E_{climb}$) with a guide to eye (dashed line).

In general, the shrinkage energy depends upon the dimensions of the simulation cell on account of the interactions amongst the periodic array of nanovoids. In the present system under study, $\Delta E_{void}$ turns out to be -4.74 eV for the loss of seven vacancies by the 2 nm diameter void. Assuming that the extra atoms enter the nanovoid one after another, the total release of shrinkage energy is found to vary nonlinearly (Fig. 5) as a function of length of the climbed segment ($l$), which fits to the equation $-\Delta E_{void} = 1.2 l^{2.47}$ eV ($l$ in nm) for the present simulations and implies unequal contribution towards the shrinkage energy from each inserted atom. Figure 5 also displays the climb energy, $E_{climb}(l)$, calculated with the method explained in section 3.1. The observation that $E_{climb}$ saturates with the climb length justifies the assumption in the analytical model of Thomson and Balluffi [2].

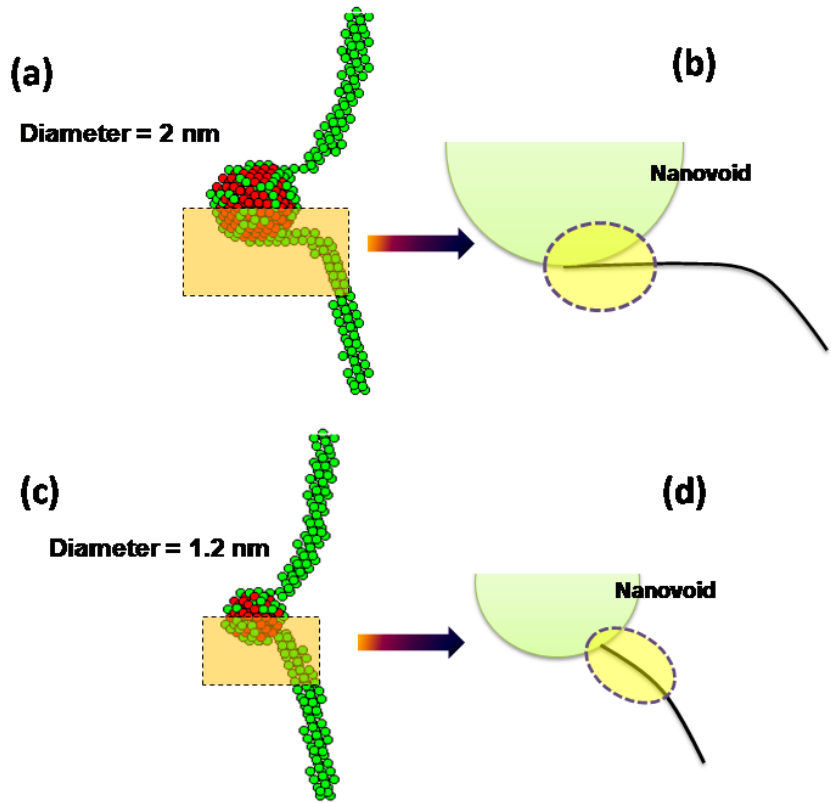

Fig. 6. Bow-out of dislocation arms during depinning at the nanovoids. (a) For the 2 nm nanovoid, large curvature of the dislocation arm causes its small inclination with the surface of the nanovoid. (b) The pinned dislocation segment has a larger field of interaction with surface atoms, which provides a barrier-free path to initiate the VIC. (c) In contrast, the curvature is small for the 1.2 nm nanovoid and (d) the field of interaction is significantly reduced. In the MD snapshots shown here, the dislocation line is observed as consisting of straight segments instead of having continuously varying curvature. This is due to the atomistic discreteness of lattice and the orientation dependent line tension, which aligns the dislocation segments along the rational crystal directions so that the line can attain an energetically favorable shape.

Figure 5 clearly shows that there exists a maximum value of $\Delta E$ ($= E_{climb} + \Delta E_{void}$), which indicates the presence of a potential energy barrier and one would expect the absence of VIC at $T = 0$ K. However, the Peach-Koehler force ($\tau b$) on the dislocation line corresponding to the applied load ($\tau$) does not have a component in the direction perpendicular to the glide plane. It implies that a suitable trajectory has been made available by the applied load such that the energy barrier to VIC is not actually encountered in this path. From the simulation snapshots (Fig. 6a), this path is recognizable as the bowing out of the pinned dislocation under shear stress. For the 2 nm nanovoid, the dislocation line has an almost screw-like orientation in the vicinity of this obstacle (Fig. 6a), similar to the observations made in the earlier reports [10,12] for nanovoids in Fe. This manifests in a larger region of interaction between the dislocation segment and the nanovoid's surface as depicted schematically in Fig. 6b. In general, the elastic displacement field along the direction of the Burgers vector of a dislocation is antisymmetric about the glide plane [1] and as a result, the atoms on the surface of the nanovoid just above the glide plane tend to move inwards while those below the glide place move outwards. In such scenario, the internal surface tension of the nanovoid comes into play, which tends to reduce this relative displacement in order to gain some reduction in the surface energy and thus, the potential gradient of the whole system drives it towards the state of dislocation climb. As we have already demonstrated that unlike the 2 nm nanovoid, the dislocation crosses the 1.2 nm diameter void with the creation of surface steps but without undergoing any climb motion (Fig. 2b), we can now perceive it in the context of the role of line curvature of the edge dislocation. Figure 6c clearly shows that that line curvature in the system with 1.2 nm diameter nanovoid is small in contrast to that for the 2 nm nanovoid shown in Fig. 6a. As a consequence, the interaction between the bowing dislocation and the nanovoid's surface becomes weak (Fig. 6d) and the potential barrier emerges, which prevents the occurrence of VIC in this case.

We can similarly analyze the issues of thermal assistance and the lowering of critical depinning load during climb. Here it can be pointed out that the potential energy barrier depicted as the maximum of $E_{climb}(l) + \Delta E_{void}(l)$ in Fig. 5 corresponds to the structural energy of the system. At elevated temperature, the critical depinning load, as well as the extent of bow-out is reduced (refer Fig. 27 in Ref. [12]), which weakens the interaction between the nanovoid and bowing out segment. In this case, an activation barrier for climb emerges which is overcome by means of thermal assistance. Although this discussion explains the role of temperature in VIC [14,16], the exact quantification of the thermal activation barrier depends upon the search of minimum energy path [22,28], which remains an open problem for further studies.

So far, we have explored the phenomenon of VIC in terms of its underlying energetics. However, it is also important to understand the structure of dislocation line during transfer of vacancies at the point of contact. It has been observed [12] that a large screw component is gained by the dislocation segment close to the nanovoid during the bow-out. How can one interpret the occurrence of climb at a point where the dislocation is almost screw-like (refer Fig. 7a)? Actually, at the moment of jog formation, a short edge type kink is formed, which undergoes the climb motion as we attempt to demonstrate in Fig. 7b. Moreover, the shapes of the two dislocation arms at the nanovoid are found to be asymmetric (Fig. 6a) due to orientation dependent line tension of the dislocation [29]. As a result, the arm which has a steeper bend develops a jog prior to the other as demonstrated in Figs. 7c and d.

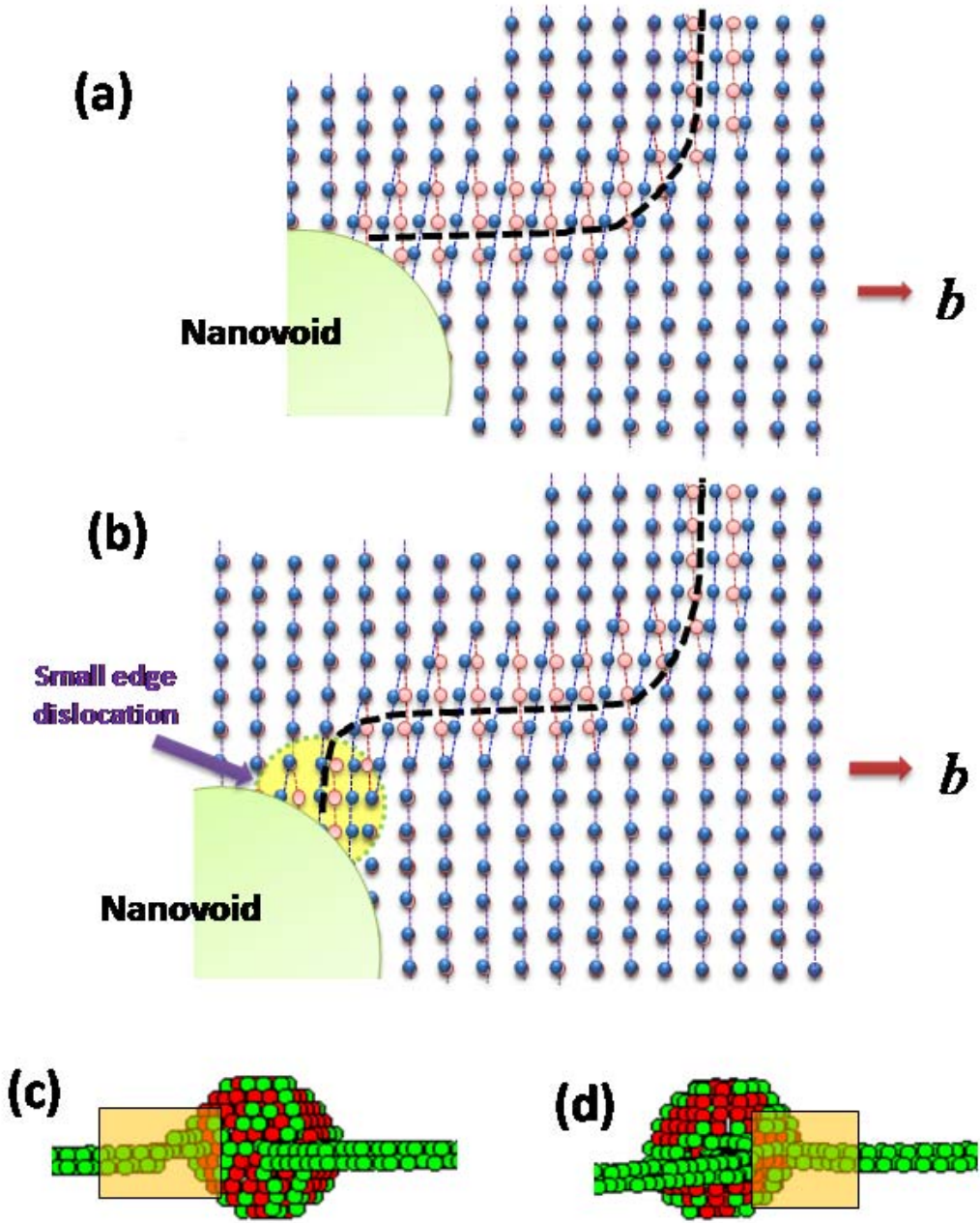

Fig. 7. (a) A simplified lattice schematic demonstrating the screw-like orientation of a dislocation arm near the nanovoid. (b) As the screw part cannot climb by absorbing vacancies, an edge type kink is formed to facilitate the VIC. (c) The jog is formed at one dislocation arm while (d) it is yet not formed in the other arm at the same instant.

## 5. Effect of VIC on the pinning strength of nanovoid

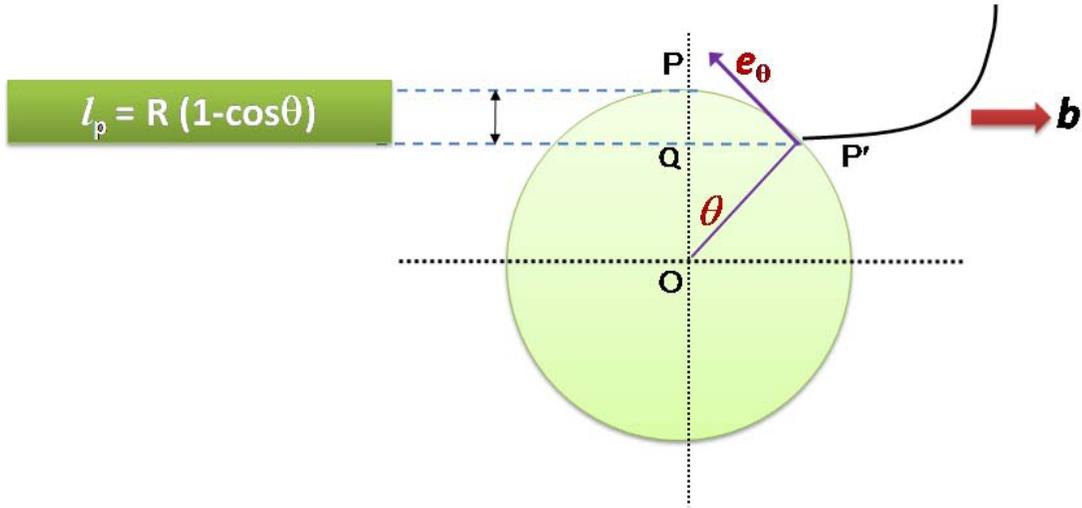

Fig. 8. $P_0$ is the point of contact between a dislocation arm and the nanovoid of radius $R$. $\theta$ is the angle of descent and the projection $PQ$ is the increment in the length of the dislocation arm that would sustain after depinning.

Next we discuss the issue of change in the depinning load due to VIC. Consider the schematic in Fig. 8, which displays a dislocation arm pinned at a nanovoid of radius R, where the point of contact ($P'$) between the dislocation and the nanovoid has descended by an angle $\theta$ from the point $P$. We can split the increment in length of the dislocation arm into two components. Firstly, the bow-out of the dislocation line due to applied load is elastic as the line curvature is only maintained by the application of external load during the state of pinning. Secondly, the projection $l_p = R(1-\cos\theta)$ corresponds to a permanent increase in the length of dislocation arm, which retains even after the breakaway. If the process of depinning occurs without the climb motion, this increment ($l_p$) can be associated with the rise in potential energy $\Delta U = E_{edge}(1-\cos\theta)$, where $E_{edge}$ is the line tension of the unbowed edge dislocation.

This leads us to introduce a resistive force at P' given by $\mathbf{F}_{res} = \frac{1}{R}\frac{d(\Delta U)}{d\theta}\mathbf{e}_\theta = 28.56\sin\theta\,\mathbf{e}_\theta$ eV/nm, during its slide along the surface of the nanovoid, where 28.56 eV/nm is the line tension of the edge dislocation in the present system, computed by the method described in section 3.2. Earlier, the elasticity theory dealt with the dislocation-nanovoid interaction in terms of the pinning force due to image stresses on the surface of the nanovoid [27,30]. It is important to note that $\mathbf{F}_{res}$, acts in addition to this image stress and is altered during the occurrence of dislocation climb. This can be incorporated by modifying $\Delta U$ to

$$\Delta U_{climb} = \Delta U + \left[E_{climb}(\theta;\theta_c) + \Delta E_{void}(\theta;\theta_c)\right]u(\theta - \theta_c), \qquad (2)$$

where $\theta_c$ is the critical angle for the onset of climb and the unit step function $u(\theta - \theta_c)$ ensures that $\Delta U_{climb} = \Delta U$ until the jog formation takes place. The resistive force at P' then becomes

$$\mathbf{F}_{res}^{climb} = \mathbf{F}_{res} + \frac{1}{R}\left[\frac{dE_{climb}}{d\theta} + \frac{d(\Delta E_{void})}{d\theta}\right]\mathbf{e}_\theta \qquad (3)$$

$$\approx \mathbf{F}_{res} + \frac{1}{R}\frac{d(\Delta E_{void})}{d\theta}\mathbf{e}_\theta, \qquad (4)$$

as $\left|\frac{dE_{climb}}{d\theta}\right| \ll \left|\frac{d(E_{void})}{d\theta}\right|$ at a later stage of climb, as evident from Fig. 5. Both the terms in Eq. (4) can be obtained from the simulations and one can observe that whenever VIC takes place, it reduces the resistive force at the point of contact and thus the critical depinning load. For instance, using the relation for $\mathbf{F}_{res}$ and the fitting relation (from Fig. 5) for $\Delta E_{void}$ obtained from our simulations with the climb length $l = R(\cos\theta_c - \cos\theta)$, Eq. (4) yields $\mathbf{F}_{res}^{climb} = \left[28.56\sin\theta - 2.96(1-\cos\theta)^{1.47}\sin\theta\right]\mathbf{e}_\theta$ eV/nm, for $\theta_c = 0°$ (which happens to be the case for one of the arms in our present simulations) and R=1 nm. Here the second term clearly indicates how the phenomenon of VIC makes $\left|\mathbf{F}_{res}^{climb}\right|<\left|\mathbf{F}_{res}\right|$, which explains the reduction in critical depinning load as reported by Osetsky *et al.* [13]. An interesting observation in this analysis is use of the results of Fig. 5, even though it does not represent the actual transition path. This is on account of the fact that the activation barrier is effective during the initiation of climb. At a later stage, the saturating tendency of $\Delta E_{climb}$ with the climb length, $l$, makes its derivative negligibly small and the basic approximation in Eq. (4) still holds. Nevertheless, an estimate of the activation profile for thermally assisted depinning would possibly yield the value of $\mathbf{F}_{res}^{climb}$ during the initiation of VIC as well.

## 6. Conclusions

The present work highlights the core physical aspects of the VIC of dislocations. In addition to describing a method of determining the line tension of dislocation from the atomistic computations, we also develop a robust simulation method to obtain the shrinkage energy of a nanovoid, which drives the process of void-induced climb. There can be more applications of this technique, including a separate study for the estimation of energies of the surface-steps on the nanovoids, which is presently underway. Besides showing the existence of energy barrier against the occurrence of VIC, we highlight the roles of dislocation's line curvature and thermal assistance to VIC. The effect of VIC on the critical depinning load of the nanovoid has also been explained. As the primary mechanism of VIC in all the systems can be understood as an interplay between the energies of jog formation and void shrinkage, the fundamental conclusions drawn from the present study remain valid for other systems with more complex climb geometries.

**Acknowledgements**

We thank Prof. David J. Bacon and Prof. G. Monnet for their valuable suggestions and remarks.

A.D. acknowledges the financial support from CSIR, India to carry out this research work.

**References**

[1] Hirth JP, Lothe J. Theory of Dislocations. New York: John Wiley and Sons; 1982.

[2] Thomson RM, Balluffi RW. J Appl Phys 1962;33:803.

[3] Balluffi RW, Thomson RM. J Appl Phys 1962;33:817.

[4] Kabir M, Lau TT, Rodney D, Yip S, Van Vliet KJ. Phys Rev Lett 2010;105:095501.

[5] MacEwen SR. J Nucl Mater 2009;54:85.

[6] Mordehai D, Martin G. Phys Rev B 2011;84:014115.

[7] Laidler JJ, Mastel B. Nature 239, 1972:97.

[8] Takeyama T, Kayano H, Takahashi H, Ohnuki S, Narui M. J Nucl Mater 1984;122-123:727.

[9] Fan Y, Kushima A, Yip S, Yildiz B. Phys Rev Lett 2011;106:125501.

[10] Osetsky YN, Bacon DJ. J Nucl Mater 2003;323:268.

[11] Osetsky YN, Bacon DJ. Mater Sci Engg A 2005;400-401:374.

[12] Bacon DJ, Osetsky YN, Rodney D. Dislocation-obstacle interactions at the atomic level, in: Hirth JP, Kubin L. (Eds.). Dislocations in Solids. Oxford: North-Holland; 2009.

[13] Osetsky YN, Bacon DJ, Mohles V. Phil Mag 2010;83:3623.

[14] Osetsky YN, Bacon DJ. Phil Mag 2010;90:945.

[15] Grammatikopoulos P, Bacon DJ, Osetsky YN. Modelling Simul Mater Sci Eng 2011;19:15004.

[16] Hatano T, Matsui H. Phys Rev B 2005;72:094105.

[17] Monnet G. Acta Mater 2007;55:5081.

[18] http://micro.stanford.edu.

[19] Finnis MW, Sinclair JE. Phil Mag A 1986; 53:161.

[20] Finnis MW, Sinclair JE. Phil Mag A 1992;50:45.

[21] Ackland GJ, Thetford R. Phil Mag A 1987;56:15.

[22] Bulatov VV, Cai W. Computer Simulations of Dislocations. Oxford: Oxford University Press; 2006.

[23] Nosé S. J Chem Phys 1984;81:511.

[24] Hoover WG, Phys Rev A 1985;31:1695.

[25] Knight DT, Burton B. Phil Mag A 1992;66:289.

[26] GZ Ganyev, Turkebaev TE. Acta Metal 1988;36:453.

[27] Scattergood RO, Bacon DJ. Acta Metal 1982;30:1665.

[28] Neria E, Fischer S, Karplus M. J Chem Phys 1996;105:1902.

[29] de Wit G, Koehler JS. Phys Rev 1959;116:1113.

[30] Haghighat SMH, Fivel MC, Fikar J, Schaeublin R. J Nucl Mater 2009;386-388:102.